# Anomalous Inner-Gap Structure in Transport Characteristics of Superconducting Junctions with Degraded Interfaces

E. Zhitlukhina[1], I. Devyatov[2,3], O. Egorov[4], M. Belogolovskii[5]*, and P. Seidel[6]

[1] *Donetsk Institute for Physics and Engineering, National Academy of Sciences of Ukraine, 03039 Kyiv, Ukraine*

[2] *Lomonosov Moscow State University, Skobeltsyn Institute of Nuclear Physics, 119991 Moscow, Russia*

[3] *Moscow Institute of Physics and Technology, Dolgoprudny, Moscow 141700, Russia*

[4] *Institut für Festkörpertheorie und -optik, Friedrich-Schiller-Universität Jena, 07743 Jena, Germany*

[5] *Institute for Metal Physics, National Academy of Sciences of Ukraine, 03680 Kyiv, Ukraine*

[6] *Institut für Festkörperphysik, Friedrich-Schiller-Universität Jena, 07743 Jena, Germany*

**Abstract**

Quantitative description of charge transport across tunneling and break-junction devices with novel superconductors encounters some problems not present, or not as severe for traditional superconducting materials. In this work, we explain unexpected features in related transport characteristics as an effect of a degraded nano-scaled sheath at the superconductor surface. Model capturing main aspects of the ballistic charge transport across hybrid superconducting structures with normally-conducting nm-thick interlayers is proposed. The calculations are based on a scattering formalism taking into account Andreev electron-into-hole (and inverse) reflections at normal metal-superconductor interfaces as well as transmission and backscattering events in insulating barriers between the electrodes. Current-voltage characteristics of such devices exhibit a rich diversity of anomalous (from the viewpoint of the standard theory) features, in particular, shift of differential-conductance maximums at gap voltages to lower positions and appearance of well-defined dips instead expected coherence peaks. We compare our results with related experimental data.

---
*Correspondence:  belogolovskii@ukr.net

# Background

The Bardeen-Cooper-Schrieffer (BCS) theory of superconductivity (S) is based on the assumption that electrons in the BCS state are bound into Cooper pairs by a sufficiently weak attractive interaction between them. These pairs are correlated so that, in order to break a pair, one has to change energies of all other pairs. It means that, unlike in the normal-metal (N) state, there is an energy gap $\Delta$ for single-particle excitations. The most complete and convincing evidence for the appearance of the energy gap came from the well-known Giaever effect of electron tunneling through N-I-S and S-I-S heterostructures where I stands for a nanometer-thick insulating layer that permits only single tunneling processes [1]. When an N-I-S trilayer with a conventional *s*-wave paring-symmetry superconductor is biased with a voltage $V$, an electron can be injected into the S side merely if its energy exceeds the gap value $\Delta_s$. It means that at the environment temperature $T \to 0$ the tunneling current $I$ across the junction is vanishing below $V_\Delta = \Delta_s/e$. The nonlinearities in $I$-$V$ curves can be more clearly seen in the quasiparticle differential conductance $dI/dV$-vs-$V$ which exhibits coherence peaks at $V = \pm V_\Delta$ for an N-I-S structure and at $V = \pm(\Delta_{s1} + \Delta_{s2})/e$ for an $S_1$-I-$S_2$ trilayer with conventional *s*-wave superconductors. This fact can be used as direct measure of the superconducting order parameter magnitude and its dependence on external parameters.

Nowadays, electron tunneling spectroscopy is a well-developed technique aimed to provide extensive information about energy spectra of conducting electrodes and an insulating interlayer between them [1]. It has been a disappointment that tunneling spectroscopy has so far given us no such evidence for novel superconductors. Our work was stimulated by systematic deviations of transport characteristics measured for nominally trilayered metal-barrier-metal heterostructures based on novel superconductors from related theoretical predictions for conventional tunnel junctions with a transmission probability $D \ll 1$ [2], as well as for S-c-S devices with $D \leq 1$ where c stands for a constriction [3].

The first finding of such kind was revealed for tunnel junctions with some transition metals, especially, for Nb-based devices and intensively discussed in 70s of the previous century long before the discovery of high-temperature superconductivity. We mean the so-called 'knee', a sudden current decrease in low-temperature *I-V* characteristics of Nb-based S-I-S junctions at voltages slightly just above a strong increase of the current that was expected to occur at the sum of $V_\Delta$ values for the two superconducting electrodes [4] (see also [5]). Moreover, the current values at very low temperatures and at voltages lower than the sum of $V_\Delta$ were considerable larger than theoretically anticipated for the BCS densities of states [6]. Later it was shown experimentally that this anomaly may be caused by the presence of a normal conducting layer on the superconducting niobium film [6,7]. This statement was confirmed theoretically [8] in the framework of the proximity-effect model which treats incoherent single-particle scatterings from a normal layer to a superconducting one within a tunnel-barrier approximation [9]. Below we discuss the origin of the 'knee' phenomenon analyzing ballistic transport in tunneling S-I-S structures where each superconducting electrode is covered by a nanometer-scale degraded sheath.

The next puzzling experimental feature was the so-called 'peak-dip-hump' structure clearly observed in conductance-vs-voltage curves for break-junctions based on high-$T_c$ BSCCO samples [10]. In these experiments, the peak was attributed to twice the gap magnitude although the values obtained were systematically lower than those provided by other experimental techniques, see Table 3.1 in [10]. But the main problem of the measurements was the presence of a pronounced dip that was revealed in data obtained by STM or ARPES [11] as well. Its derivation has not been clear and, in particular, was believed to represent a physical quantity responsible for pairing in the studied compound [12]. It was suggested [13] that the dip appears at the voltage bias $(2\Delta+\Omega)/e$ where $\Omega$ is the energy of some collective excitations that performs a role of glue for paired electrons. The author of ref. [10] studied how the estimated doping level influences the peak and observed a large spread of the measured peak positions. Below we reproduce numerically the 'peak-dip-hump' structure and propose our explanation of each of the three elements relating them to the presence of a degraded layer on the surface of the high-$T_c$ superconductor. It should be noticed that at early stages of single-electron tunneling experiments four-layered N-I-n-S heterostructures with a normal (n) interlayer between an ordinary superconductor and the barrier were fabricated and studied. Comparing related data with our simulations, we show that, at least qualitatively, the measured conductance spectra of S-I-n-S samples having the n interlayer for sure do follow our predictions.

The last unusual feature which will be analyzed in the paper relates the inner-gap structure in S-c-S junctions where c stands for a constriction with a transmission probability $D \leq 1$. The authors of ref. [14] fabricated two types of symmetric S-c-S trilayers, Al-/InAs-nanowire-Al and Nb-InAs-nanowire-Nb junctions and measured the differential resistance d$V$/d$I$ as a function of the voltage bias at various temperatures. At 0.4 K, the Al-based devices exhibited subharmonic gap features at $V_n = 2\Delta_{Al}/(en)$, $n$ is an integer. Due to the theory [3], such features are the manifestation of multiple Andreev reflections within the constriction between two identical superconductors and they should expose themselves as local minimums in the resistance-vs-voltage dependence as it was indeed observed for Al-based samples, Fig. 1a in ref. [14]. Similar $dV/dI$ measurements at $T < 4$ K for a typical Nb-based junction depicted in Fig. 1b [14] revealed a subharmonic gap structure corresponding to $2\Delta_{Nb}/e$ and $\Delta_{Nb}/e$ but they were **peaks rather than expected dips**. Moreover, it was not the only observation of the difference between the two types of junctions. As was stressed in ref. [14], see related references therein, in the literature there is a large number of findings with resistance peaks at $V_n = 2\Delta_{Nb}/(en)$ in d$V$/d$I$ curves for Nb-based S-c-S junctions and at the same time dips in d$V$/d$I$ curves for Al-based structures. Below we explain the discrepancy as a result of the degradation of Nb-c and c-Nb interfaces.

Our main results which throw some light on unexpected features of transport characteristics of superconducting heterostructures with degraded interfaces are summarized in the conclusions.

**Methods**

**General considerations**

We start with a model N-I-n-S system which consists of a normal counter-electrode, a nm-thick insulating barrier, a superconducting electrode studied, and a nano-scale clean non-superconducting (n) interlayer between I and S films. Confinement of electrons in the n-metal film results in discrete quantum-well states that can be probed directly by single-electron tunneling spectroscopy. Below we provide an intuitive geometric picture for the localized resonances that play decisive role in the formation of related transport characteristics. The important impact of the bound states on the charge transport across S-N-S Josephson junctions and ability to probe them by tunneling experiments are well known (see the review [15]). In this work, we show that some unusual features revealed in novel superconductors by single-electron tunneling and break-junction techniques can be understood as an effect of the bound states within a degraded nano-scale sheath at the superconductor surface.

To calculate energies $E$ of the bound states in ballistic junctions we should take into account the principal difference between backscattering processes at I-n and n-S interfaces. In the first case, an electron (hole) incident on the I-n interface from the n side is retroreflected into an electron (hole) of the same energy $E$ and of the same absolute value of momentum but travelling in the opposite direction to the incoming charge. On the contrary, the reflected from the n-S interface quasiparticle has the same energy $E$ and almost the same momentum is travelling in the opposite direction, and its charge is opposite in sign to that of the incident quasiparticle. Such process is known as Andreev reflection [16], an elastic quasielectron-into-quasihole transformation of Bogoliubov quasiparticles and inverse (with a missing charge of 2e absorbed into the superconducting ground state as a Cooper pair).

Taking it into account, we can easily understand the origin of Andreev bound states within the energy gap that are formed in the n interlayer of the thickness $d_n$ and find their energies $\bar{E}$ from the demand of coherent superposition of reflected from I-n and n-S interfaces quasiparticle waves. An electron "round-trip" inside the n interlayer consists of four scattering processes, an Andreev electron-into-hole transformation at the n-S interface, a specular hole-into-hole reflection at the I-n boundary, a hole-into-electron transformation at the n-S interface, and an electron-into-electron reflection at the I-n boundary, and four passages across the n interlayer, two of them as a quasielectron and two of them as a quasihole with the phases accumulated $\varphi^e = k_x^e d_n$ and $\varphi^h = -k_x^h d_n$, respectively; $\mathbf{k}^{e(h)}$ is an electron (hole) wave vector, note that a hole is moving in the direction opposite to that of its wave vector. The reflection coefficients can be calculated from the boundary conditions for the wave functions. Each backscattering from the insulating layer provides a phase shift of $\pi$ whereas the Andreev reflection within the energy gap of an *s*-wave superconductor contributes an additional phase shift $\chi^{eh(he)}(\varepsilon) = -\arccos(\varepsilon/\Delta_s)$, $\varepsilon = E - E_F$, $E_F$ is the Fermi energy. Adding the phases accumulated along an electron 'round-trip' in the n interlayer with two subsequent Andreev reflections, we get the following expression for the phase shift

$$\Delta\varphi_s = k_x^e d_n + \chi^{eh} - k_x^h d_n + \pi - k_x^h d_n + \chi^{he} + k_x^e d_n + \pi = \frac{4\varepsilon}{\hbar v_F} \frac{d_n}{\cos\theta} - 2\arccos(\varepsilon/\Delta_s) + 2\pi \quad (1)$$

which is valid for an *s*-wave superconducting electrode, θ is the incident angle, $v_F$ is the Fermi velocity.

The Bohr-Sommerfeld quantization rule requires an electron wave-function phase accumulation $\Delta\varphi_s$ along an enclosed propagating loop inside the n interlayer to be an integer multiple of 2π. Thus, the lowest bound level $\bar{\varepsilon} = \bar{E} - E_F$ follows from the relation

$$\bar{\varepsilon}(\theta) = (\hbar v_F / 2d_n) \arccos(\bar{\varepsilon}(\theta) / \Delta_s) \cos\theta \qquad (2)$$

valid when $d_n \neq 0$. For vanishing $d_n$ we get the classical result $\bar{\varepsilon} = \Delta_s$, i.e., a well pronounced singularity at the energy gap value. When the thickness $d_n$ is finite, $\bar{\varepsilon} < \Delta_s$. It is clear from Eq. (2) that its effect on current-voltage characteristics is determined by the ratio $\alpha = 2d_n \Delta_s / (\hbar v_F) = k_F d_n (\Delta_s / E_F) = d_n / \xi_n$ where $\xi_n$ is the relevant length scale which is determined by electronic characteristics of the n-metal and the gap value of the S-electrode in contact. In the most traditional superconductors, like Pb, Sn, etc., due to comparatively large $v_F$ and small $\Delta_s$ $\xi_n \geq 100$ nm and, hence, the effect of the n-interlayer can be observed merely for $d_n$ of the order of tens nanometers. Such interlayers can be introduced within the junction only artificially. If however the Fermi velocity is small and at the same time the energy gap is large, as it is in transition metals and novel high-$T_c$ compounds, a completely different situation is expected. In this case, degradation of the superconductor surface on the length scale of the order of several nanometers which is typical for these materials can result in crucial modifications in related transport characteristics. We shall discuss these changes and the information about the S-electrode spectra that can be extracted from such measurements.

**Charge scattering characteristics**

In the following, we approximate the spectrum of electrons in all conducting layers by parabolic bands and limit ourselves to the planar geometry of the studied heterostructures with an x-axis normal to interfaces. Next, we assume that the electron wave function may be factorized into in-plane and tunneling-direction components and will discuss only the latter one considering the in-plane momentum $\mathbf{k}_\parallel$ constant. For a non-superconducting metallic interlayer it reads as $\psi_\perp^{(n)}(x) \sim \exp(ik_x^{(n)}x - x/(2l_n))$, where $\mathbf{k}^{(n)} = (k_x^{(n)}, \mathbf{k}_\parallel)$ is the wave vector of an excitation with the energy ε, which is a solution of the equation $k_x^{(n)} = \sqrt{2m(E_F \pm \varepsilon)/\hbar^2 - \mathbf{k}_\parallel^2}$, the sign ± corresponds to electron (e) and hole (h) excitations, respectively, $E_F$ is the Fermi energy, $m$ is the quasiparticle mass. To take into account the finite value of the mean free path of a quasiparticle excitation $l_n$ in a nm-thick n-interlayer we have introduced an additional imaginary term $\pm i/(2l_n)$ in the wave vector $k_x^{(n)}$ of an electron (a hole) in the n interlayer, see also [17]. Last, for the sake of simplicity, we assume that Fermi energies in the conducting layers, including the superconducting one, are the same.

Notice that the stepwise approximation for the superconducting pair potential in the n-S bilayer used below and known as a rigid-boundary condition is not self-consistent. According to Likharev [18], deviation of the self-consistent solution from the step-like function strongly decreases when the interface resistivity is much bigger than that of metal electrodes. It is very difficult to estimate the thickness of the degraded surface layer as well as the transparency of its interface with the bulk. Thus our results can only claim a qualitative explanation of the anomalous in-gap features.

To get analytical expressions for elementary scattering amplitudes in the N-I-n-S structure, we model the barrier I by a short-range repulsive potential of a rectangular shape $U(x) = U_0$ and the thickness $d_I$. The transparency of such barrier equals to $D_I = \left[1 + \left(k_F^2 + \kappa^2\right)^2 \text{sh}^2 \kappa d_I / (4k_F^2 \kappa^2)\right]^{-1}$ [19] where $\kappa^{-1} = \hbar/\sqrt{2m(U_0 - \varepsilon)}$ is the decaying depth of a quasiparticle wave function. When $\kappa d_I \ll 1$ it reads $D_I = (1 + (k_F^2 + \kappa^2)^2 d_I^2 / (4k_F^2))^{-1}$. Introducing a dimensional parameter $Z = (k_F^2 + \kappa^2) d_I / (2k_F)$, we get a simple Lorentzian $D_I = 1/(1 + Z^2)$ (in the limit $\kappa \gg k_F$ it coincides with related results of the paper [2]). Within the same $\kappa d_1 \ll 1$ approximation we can derive reflection and transmission probability amplitudes for an electron (hole) to be transmitted through the barrier

$$r^e(\theta) = (r^h(\theta))^* = -Z/(Z - i\cos\theta), \quad t^e(\theta) = (t^h(\theta))^* = -i\cos\theta/(Z - i\cos\theta).$$

The elastic scattering process by which normal current is transferred to a supercurrent at the interface between a superconductor and a non-superconducting metal, the Andreev effect [16], lies in the fact that an electron (hole) incident on the interface from the normal side is retroreflected into a hole (electron). In the general case the related $r^{eh}$ (an electron is scattered into a hole) and $r^{he}$ (a hole-electron transformation) scattering characteristics look as in [20]

$$r^{eh(he)}(\theta) = \frac{\varepsilon - h(\varepsilon)}{|\Delta(\theta)|} \exp(\mp i\Phi(\theta)), \tag{3}$$

where $h(\varepsilon) = \text{sign}(\varepsilon)\sqrt{\varepsilon^2 - |\Delta(\theta)|^2}$ for $|\varepsilon| > \Delta(\theta)$ and $h(\varepsilon) = i\sqrt{|\Delta(\theta)|^2 - \varepsilon^2}$ for $|\varepsilon| < \Delta(\theta)$. Here $\Delta(\theta)$ is an order parameter which is constant $\Delta_s$ for s-wave pairing realized in all conventional superconductors, $\Phi(\theta)$ is the order parameter phase. In high-$T_c$ compounds it is widely accepted that such superconductors have a $d_{x_a^2-x_b^2}$-wave pairing symmetry. The energy gap of such a pairing state manifests a sign change at some directions of the Fermi wave vector and its angle dependence is $\Delta(\theta) = \Delta_d \cos[2(\theta - \gamma)]$ with $\gamma$, the misorientation angle between the surface normal and the crystalline axis along which the order parameter reaches maximum.

**Results and discussion**

**Transport characteristics of N-I-n-S junctions**

We start with a simplest case of superconducting heterostructures with normal interlayers. To calculate the current $I$-vs-voltage $V$ curves for N-I-n-S junctions with an *s*-wave superconductor, we use the Landauer-Büttiker formalism applied to superconductor-based structures [21]

$$I(V) = \frac{1}{eR_N} \int_{-E_F}^{\infty} d\varepsilon [f(\varepsilon - eV) - f(\varepsilon)] \int \frac{d^2 k_{\parallel}}{(2\pi)^2} D(\varepsilon, \theta), \quad (4)$$

where $R_N$ is the normal-state resistance, $\theta$ is the injection angle between an electron wave vector and the *x* axis, the reference potential of a superconducting side is put to zero, $f(\varepsilon)$ is the Fermi-Dirac distribution function, $D(\varepsilon, \theta)$ is the electron penetration probability. It is easier to calculate the latter quantity in an N part of the structure where

$$D(\varepsilon, \theta) = 1 - |R^{ee}(\varepsilon, \theta)|^2 + |R^{eh}(\varepsilon, \theta)|^2, \quad (5)$$

$R^{ee}(\varepsilon, \theta)$ and $R^{eh}(\varepsilon, \theta)$ are angle-dependent probability amplitudes for an electron entering the N-electrode to be scattered back as an electron or as a hole, respectively. To obtain them, we interpret the charge transmission across a heterostructure as a sequence of an infinite number of interface scattering events including Andreev electron-hole and vice versa transformations at N/S boundaries [22,23]. Then we get

$$R^{eh} = t^e e^{i\varphi^e} r^{eh} e^{i\varphi^h} t^h (1 + r^h e^{i\varphi^h} r^{he} e^{i\varphi^e} r^e e^{i\varphi^e} r^{eh} e^{i\varphi^h} + \ldots) = \frac{t^e e^{i\varphi^e + i\varphi^h} r^{eh} t^h}{1 - r^h e^{2i\varphi^e + 2i\varphi^h} r^{he} r^{eh} r'^e};$$

(6)

$$R^{ee} = r^e + t^e e^{i\varphi^e} r^{eh} e^{i\varphi^h} r^h e^{i\varphi^h} r^{he} e^{i\varphi^e} t^e (1 + r^h e^{i\varphi^h} r^{he} e^{i\varphi^e} r^e e^{i\varphi^e} r^{eh} e^{i\varphi^h} + \ldots) =$$

$$= r^e + \frac{t^e e^{2i\varphi^e + 2i\varphi^h} r^{eh} r^h r^{he} t^e}{1 - r^h e^{2i\varphi^e + 2i\varphi^h} r^{he} r^{eh} r'^e},$$

where $\varphi^{e(h)} = \pm k_x^{e(h)} d_n + i d_n / (2 l_n)$ is the complex-valued phase shift acquired during an electron (hole) path from one edge of the n-interlayer to the other. Eqs. (6) provide an insight into an effect of the non-superconducting interlayer on transport characteristics of N-I-n-S junctions. Note that formally it is reduced to multiplication of the standard formula (3) for the Andreev scattering amplitude by a factor $\exp(i\varphi^e + i\varphi^h)$. Such procedure permits to use all previously developed expressions for N-I-S and S-I-S junctions by only modifying related Andreev scattering coefficients.

Fig. 1 exhibits differential-conductance curves for a planar three-dimensional N-I-n-S structure with an *s*-wave superconductor and a normal n interlayer where the probability to transfer the barrier I is much less than unity (the tunneling regime). The impact of the n-layer thickness $d_n$ and that of the mean free path $l_n$ are controlled by the parameters $\alpha = 2d_n\Delta_s/(\hbar v_F)$ and $\beta = d_n/l_n$, respectively. In the limit $d_n \to 0$ we get a well-known coherence peak in the differential conductance curve at $V = V_\Delta$ (Fig. 1, solid curve) that reflects the presence of a surface bound state in the n interlayer of a vanishing thickness, i.e., at the IS interface. Its asymmetry arises due to the non-analytical behavior of the scattering amplitudes $r^{eh(he)}(\varepsilon)$ at $\varepsilon = eV_\Delta$. With increasing $d_n$, the bound state formed due to the interference of electron and hole waves is shifted to lower voltages and a pronounced dip appears slightly below $V = V_\Delta$ (Fig. 1, dashed curve). For larger $d_n$ we can see the appearance of an additional hump structure (Fig. 1, dotted curve) above $V = V_\Delta$ which reminds about the presence of a non-analytical square-root dependence of Andreev-reflection amplitudes (3). All three main elements of the peak-dip-hump structure are shown by arrows in the inset in Fig. 1 that illustrates the effect of the electron mean free path on the conductance spectra of the tunneling N-I-n-S(*s*-wave) structure.

Fig. 2 shows that the presence of a peak-dip-hump structure does not strongly depend on the barrier transparency. It is more pronounced in the tunneling regime ($D_I \ll 1$) but is well reproduced even for a high-transparent device with $D_I = 0.5$. As is expected, the temperature and charge scattering processes significantly destroy the anomalous features (see the insets in Figs 1 and 2).

Before turning to novel high-temperature superconductors, let us discuss old experiments on four-layered N-I-n-S structures having the n interlayer for sure. Here we refer to three papers with n = Cu and S = Pb [24.25] as well as with n = Al and S = NbZr alloy [26]. In the first paper [24], the authors studied Cu-Pb sandwiches with a 500 nm-thick lead film and copper interlayers of different thicknesses. At 0.06 K they observed a sharp dip at about the lead energy gap for $d_{Cu}$ between 40 and 120 nm, for higher thicknesses the amplitude of the singularity was vanishing, probably, due to the impurity scattering inside the Cu interlayer [24]. These results were confirmed in ref. [25] for $d_{Cu}$ = 35 and 75 nm and $d_{Pb}$ = 700 nm. The authors [24,25] tried to interpret their data within the McMillan tunneling model of the superconducting proximity effect [9] but experienced large discrepancies, in particular, in the dip region. To explain them, the authors of ref. [24] supposed that the dip is the result of quasiparticle interference in the copper interlayer. Our simulations, Figs. 1 and 2, are based just on this assumption and qualitatively agree with the data shown in [24-26]. Measurements of In-I-Al-NbZr alloy junctions [26] also revealed the peak-dip structure that was interpreted as a bound state within the Al interlayer. Notice that additional peaks and dips in the differential conductance of tunnel junctions in ref. [26] can arise due to resonant tunneling processes across localized states in the interlayer [27].

Further confirmation of the bound-state assertion for the conductance peak came from an experiment [28] with a ferromagnetic (F) nm-thick film which replaced the normal interlayer in the four-layered junction. In this case the charge reflected at the F-S interface is created in the electron density of states with a spin opposite to that of the incident quasiparticle and Eq. (1) is modified as follows

$$\Delta\varphi_s = k_{Fx}^{\uparrow}d_F + \chi^{eh} - k_{Fx}^{\downarrow}d_F + \pi - k_{Fx}^{\downarrow}d_F + \chi^{he} + k_{Fx}^{\uparrow}d_F + \pi = 2\delta k_x^{(F)}d_F - 2\arccos(\bar{\varepsilon}/\Delta_S) = 2\pi j$$

Since, for example, the difference in wave vectors $\delta k^{(F)} = k_F^{\uparrow} - k_F^{\downarrow}$ in Ni is of the order of several nm$^{-1}$, effects very similar to those predicted for N-I-n-S junctions with $d_n$ of the order of tens nanometers can be now observed in N-I-F-S ones with a few nm thick Ni interlayer. Indeed, an anomalous ''double-peak'' conductance feature revealed in ref. [28] at $V$ near $\Delta_{Nb}/e$, in our opinion, is a peak-dip-hump structure with a maximum strongly smeared by local inhomogeneities within the F interlayer.

We should note that the calculations shown in Fig. 1 relate to a conventional $s$–wave superconductor while the peak-dip-hump structure was revealed in high-$T_c$ cuprates with $d$-wave pairing. Thus we should now discuss the latter case where the order parameter is an angle function $\Delta(\theta) = \Delta_d \cos[2(\theta - \gamma)]$. If the misorientation angle $\gamma = 0$, we have the same phase shift (1) as in the conventional $s$-wave superconductor with the only exception, the angle dependence of the gap parameter $\tilde{\Delta}(\theta) = \Delta_d \cos(2\theta)$. The conductance spectrum for such an orientation is shown in Fig. 3. As in the case of an $s$-wave superconductor, it exhibits a peak-dip-hump structure for sufficiently thick normal interlayers. On the contrary, if $\gamma = \pi/4$, we get the following expression for the angle-dependent phase shift

$$\Delta\varphi_d = \frac{4\varepsilon}{\hbar v_F} \frac{d_n}{\cos\theta} - 2\arccos\left(\frac{\varepsilon}{\Delta_d |\sin(2\theta)|}\right) - \pi + 2\pi \ , \qquad (7)$$

It is important that for such film orientation, independently on the n-layer thickness $d_n$ the phase shift $\Delta\varphi_d$ vanishes at $\varepsilon = 0$. Hence, in this case we should always observe a zero-bias peak which is known to be a $d$-wave superconductivity indicator [29]. For comparatively thick interlayers, in addition to the zero-bias anomaly, we register, at first, a small peak at voltage bias $V_\Delta$, and with the further increase of $d_n$ the emergence of a peak-dip-hump structure, see Fig. 4.

Since high-$T_c$ samples are often composites of randomly oriented superconducting grains, we predict that, in the presence of an additional n interlayer, their conductance spectra should exhibit a pronounced peak-dip-hump structure which is present in all orientations and a weak zero-bias anomaly specific for a certain tunneling direction.

**Transport characteristics of S-n-I-n-S junctions**

Physical mechanisms of the charge transmission across heterostructures with two superconducting electrodes are more complicated. In this section we limit ourselves to *s*-wave pairing taking into account the fact that, after averaging over the realizations of the disorder, the order parameter in composites of randomly oriented superconducting grains with *d*-wave paring has global s-wave symmetry [30].

Let us start with simplest tunneling transmission processes in S-n-I-n-S junctions. The presence of bound states at the electrode surfaces gives rise to unusual features in transport characteristics when the energies of the discrete levels align by applying the external bias. When properly aligned, we get the tunneling current peak and a negative differential resistance just above the resonant bias. Our calculations were done using a standard formula for tunneling current in a symmetric S-n-I-n-S configuration [1] where the tunneling density of states should be replaced by the electron penetration probability (5). The results are shown in Fig. 5. Calculated current-voltage characteristic for a three-dimensional S-n-I-n-S junction well reproduces the 'knee' feature, see the section 'Background'. Moreover, the first derivative shown in the inset in Fig. 5 demonstrates a peak, a region of the negative differential conductance which can reveal itself as a dip, and a hump at $V = 2\Delta_s/e$. Note that just the position of the hump but not that of the peak determines the doubled value of the energy gap $\Delta_s$ in the studied *s*–wave superconductor.

Now let us discuss a high-transparency junction formed by two superconductors with degraded surfaces. When the length of the transition region between the two superconducting electrodes is less than elastic and inelastic lengths, we can again describe the transport across the device in terms of Andreev-reflection amplitudes and barrier scattering characteristics. But in this case, comparing to the tunneling regime, a new specific feature largely complicates the calculations. Let us look at an electron-like quasiparticle injected, for example, from the left electrode. Its energy after transferring the barrier is increased by $eV$. The Andreev-reflected hole-like excitation is moving in the opposite direction to that of the injected electron and since of the opposite charge sign, its energy also increases by $eV$ after the transmission across the barrier. These scattering events will continue back and forth in the normal interspace between the superconductors and, as a result, each round trip of an electron-like quasiparticle will increase its energy by the value of $2eV$. Finally, from each side of the barrier we get an infinite set of scattering states with different energies shifted by $2eV$. It results in recurrence relations for amplitudes of electron-like and hole-like wave functions which have been solved numerically. Our numerical simulations at zero temperature repeated in general terms similar calculations developed earlier [20,31-33] with the only exception, the presence of an additional phase shift originated from charge passages across the n-interlayer. Due to the standard theory of multiple Andreev reflections, the peaks positions in the conductance spectra are determined by the formula $V_n = 2\Delta_s / n$ where $n$ is an integer. Account of the additional phase shifts in the degraded near-surface regions replaces the peaks by dips (see Fig. 6). It is just what was found in ref. [14] and some previous publications for related Nb-based samples. The difference between Al-based junctions that follow the conventional theory [3] and Nb-based structures which contradict it arises from the existence of a proximity layer, either normal, or superconducting with reduced critical temperature, at the surface of a superconducting Nb film whereas the Al layer is usually not spoiled.

**Conclusions**

We have presented scattering-like approach for studying Andreev bound states in a normal interlayer within ballistic superconducting heterostructures. Calculated transport characteristics of such devices exhibit a rich diversity of anomalous (from the viewpoint of the standard theory) transport characteristics, in particular, shift of differential-conductance maxima at gap voltages to lower positions and appearance of well-defined dips instead expected coherence peaks. We have compared our results with related experimental data and explained unexpected features observed in transport characteristics of heterostructures with transition metals and novel superconductors as an effect of the additional non-superconducting interlayer at their surfaces. Note that our results are qualitatively similar to those obtained by other authors in the framework of physical assumptions that significantly differ from ours [8,34,35].

**Competing interests**

The authors declare that they have no competing interests.

**Acknowledgements**

This work was performed within the German-Ukrainian project SE 664/18-1 supported by Deutsche Forschungsgemeinschaft (DFG) and within the grant No. 612600 LIMACONA "Light-Matter Coupling in Composite Nano-Structures" supported by the EU Seventh Framework Programme, I.D. acknowledges financial support from the Russian Foundation for Basic Research, projects 13-02-01085, 15-52-50054 and financial support from the Ministry of Education and Science of the Russian Federation, contract 14.Y26.31.0007.

# FIGURES

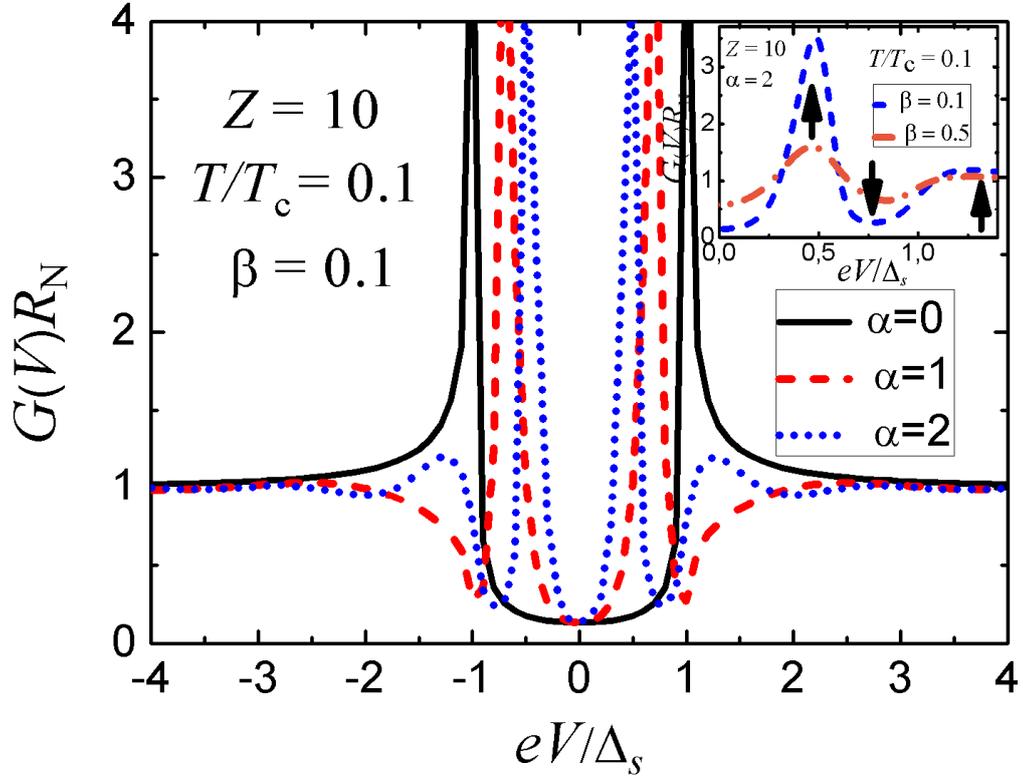

**Fig. 1** Main panel: Differential conductance-versus-voltage characteristics for a planar three-dimensional N-I-n-S(*s*-wave) junction with various thicknesses of the normal n interlayer in the tunneling regime ($D_I \ll 1$); parameters $\alpha = 2 d_n \Delta_s / (\hbar v_F)$ and $\beta = d_n / l_n$. Inset: Effect of the electron mean free path on the conductance spectra of a tunneling N-I-n-S(*s*-wave) four-layered device; the three arrows show the main elements of the peak-dip-hump structure discussed in the text

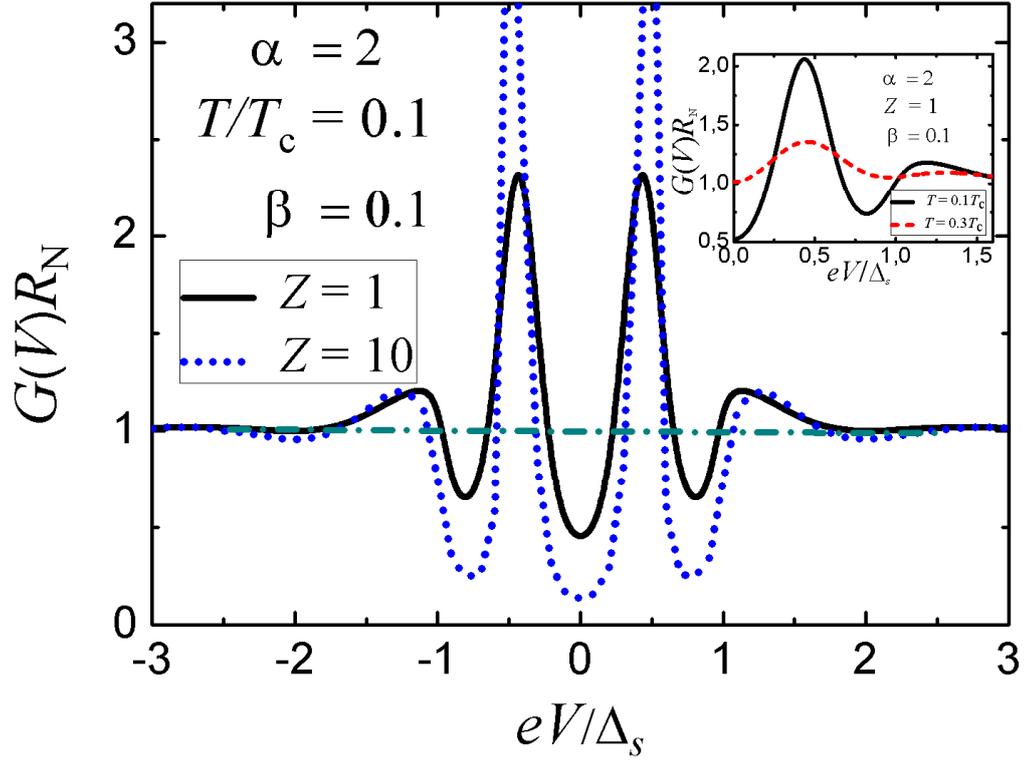

**Fig. 2** Main panel: Effect of the barrier transparency on the peak-dip-hump structure in conductance spectra of a planar three-dimensional N-I-n-S(*s*-wave) junction; parameters $\alpha = 2d_n\Delta_s/(\hbar v_F)$ and $\beta = d_n/l_n$. Inset: Temperature effect on differential conductance-versus-voltage characteristics for a planar three-dimensional N-I-n-S structure with a high-transparency tunneling barrier ($D_I = 0.5$)

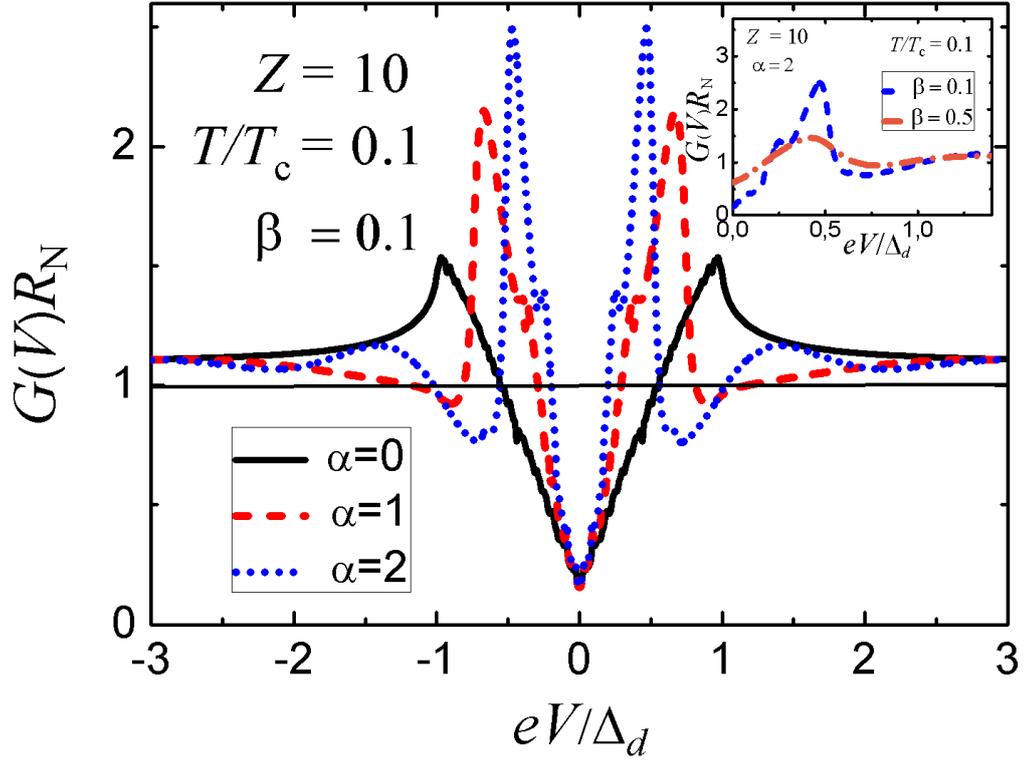

**Fig. 3** Main panel: Differential conductance-versus-voltage characteristics for a planar three-dimensional N-I-n-S(*d*-wave) structure with different thicknesses of the normal n interlayer in the tunneling regime ($D_I \ll 1$); the angle $\gamma = 0$; parameters $\alpha = 2d_n\Delta_d/(\hbar v_F)$ and $\beta = d_n/l_n$. Inset: Effect of the electron mean free path on the conductance spectra of the tunneling N-I-n-S(*d*-wave) structure

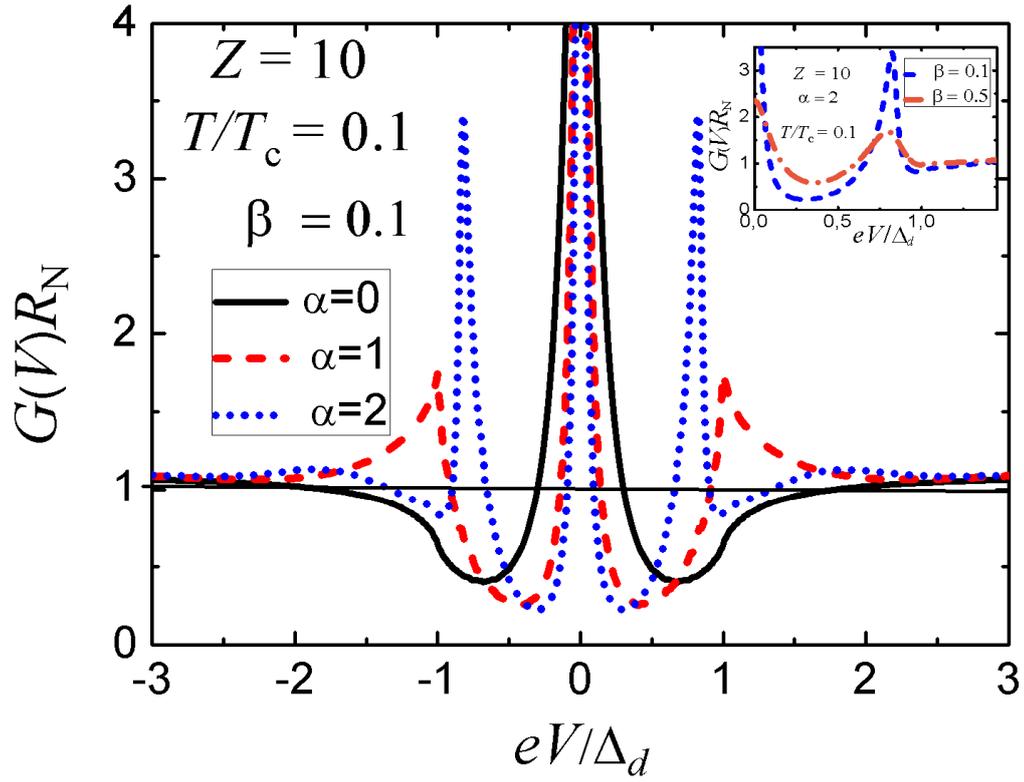

**Fig. 4** Main panel: Differential conductance-versus-voltage characteristics for a planar three-dimensional N-I-n-S($d$-wave) structure with different thicknesses of the normal n interlayer in the tunneling regime ($D_I \ll 1$); the angle $\gamma=45°$; parameters $\alpha = 2d_n\Delta_d/(\hbar v_F)$ and $\beta = d_n/l_n$ Inset: Effect of the electron mean free path on the conductance spectra of a tunneling N-I-n-S($d$-wave) structure

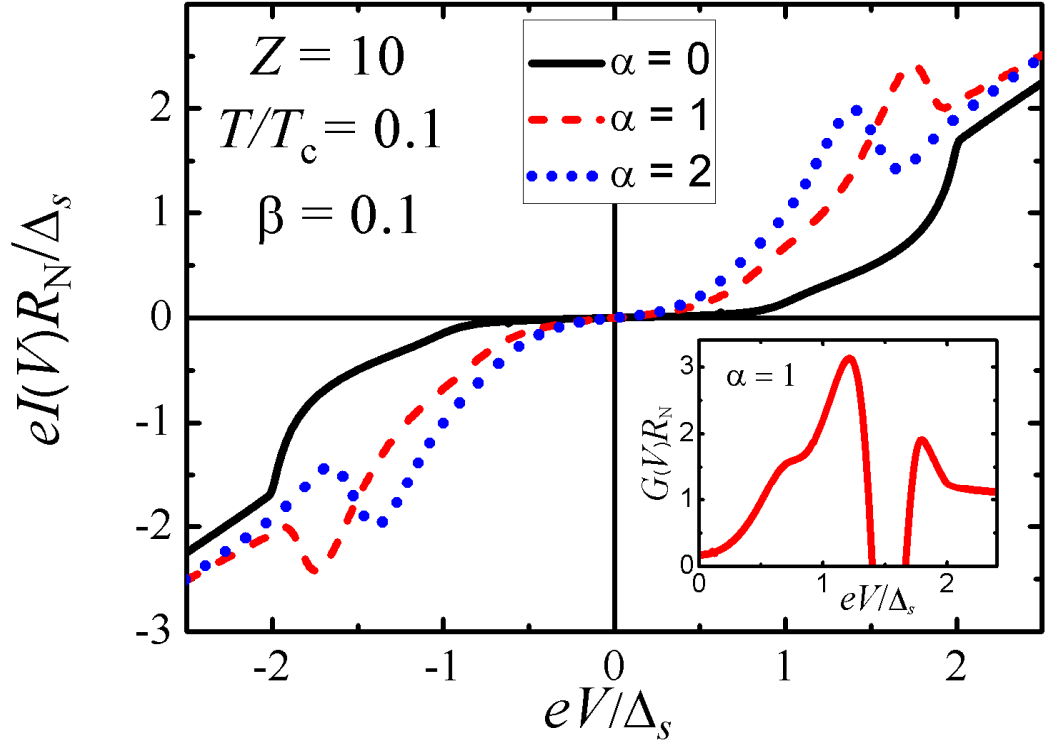

**Fig. 5** Main panel: Current-voltage characteristics for a planar three-dimensional S-n-I-n-S junction with identical *s*-wave superconductors and various thicknesses of the normal n interlayer in the tunneling regime ($D_I \ll 1$). Inset: Conductance spectra of the S-n-I-n-S device, $\alpha = 1$

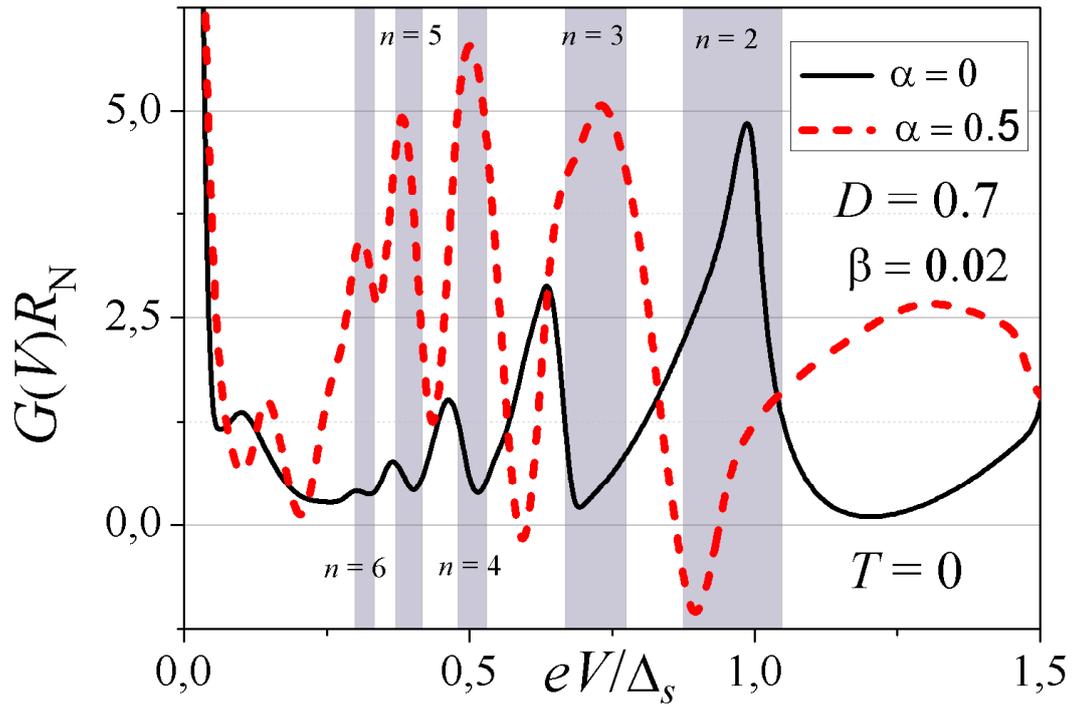

**Fig. 6** Differential conductance-versus-voltage spectra for a planar S-n-I-n-S junction with identical *s*-wave superconductors and high-transparency transition region ($D_I = 0.7$) without n-interlayer ($\alpha = 0$) and for a finite thickness of the normal n interlayer. Shaded regions show positions of the $2\Delta_s/n$ features (peaks in the first case and dips in the second case) in the conductance spectra